\documentstyle[12pt]{article}     % Paper without sections

\def\monthyear{\ifcase\month\or
  January\or February\or March\or April\or May\or June\or
  July\or August\or September\or October\or November\or
December\fi
  \space\number\year}

\def\up#1{\leavevmode \raise.16ex\hbox{#1}}

\def\slash#1{{\mathpalette\c@ncel{#1}}}

\renewcommand{\baselinestretch}{1.17}
\newcommand{\gapproxeq}{\lower
.7ex\hbox{$\;\stackrel{\textstyle >}{\sim}\;$}}
\newcommand{\lapproxeq}{\lower
.7ex\hbox{$\;\stackrel{\textstyle <}{\sim}\;$}}

\newcounter{appendice}

\def\thefiglist#1{\section*{Figure Captions\markboth
 {FIGURE CAPTIONS}{FIGURE CAPTIONS}}\list
 {Figure \arabic{enumi}.}
 {\settowidth\labelwidth{Figure #1.}\leftmargin\labelwidth
 \advance\leftmargin\labelsep
 \usecounter{enumi}}
 \def\baselinestretch{1.1}\@normalsize
 \def\newblock{\hskip .11em plus .33em minus -.07em}
 \sloppy}

\newcommand{\be}{\begin{equation}}
\newcommand{\ee}{\end{equation}}
\newcommand{\bea}{\begin{eqnarray}}
\newcommand{\eea}{\end{eqnarray}}
\newcommand{\bean}{\begin{eqnarray*}}
\newcommand{\eean}{\end{eqnarray*}}

\begin{document}
\begin{titlepage}
\begin{flushright} RAL-93-034\\
\end{flushright}
\vskip 2cm
\begin{center}
{\bf\large The Nucleon Spin Crisis Bible}
\vskip 2cm
 {\bf \large F E Close\\
Rutherford Appleton Laboratory,\\
Chilton, Didcot, Oxon,
OX11 OQX, England.}\\
\end{center}
\begin{abstract}
When the new data on polarised lepton nucleon scattering are compared at the
same value of $Q^2$ and with a common set of assumptions, a consistent picture
of the spin content of the nucleon begins to emerge.  Higher order effects in
$0(\alpha_s)$, higher twist effects, modern data on unpolarised structure
functions and an updated value for F/D are all important in analysing the data.

The detailed x dependences of the polarisation asymmetry in the valence quark
region are shown to confirm 20 year old predictions of the quark model and I
argue that these are an important ingredient in decoding the nucleon spin
puzzle.
\end{abstract}
\vspace*{3 in}
Invited talk at the 6th ICTP Workshop, Trieste \\May 3-7, 1993
\end{titlepage}

Deep inelastic scattering of polarised leptons off polarised nucleons reveals
th
 e
internal spin structure of the target.  In 1988 the EMC results for such an
experiment
generated considerable excitement as their data seemed to imply that very
little

of a
proton's spin is due to its quark constituents [1,2].  This generated
considerab
 le
activity both by theorists and experimentalists.  Theorists were inspired to
loo
 k
more
deeply into the foundations of polarised inelastic formalism; the role of the
anomaly,
polarised gluons and polarised strangeness within the proton have all been
intensively examined.  On the experimental side, a round of second generation
experiments was planned and, just weeks ago, we learned of the first data
on
neutron polarisation.  It is for this reason that the question of the nucleon's
 spin
has
received the lead attention at this conference.

The new experimental data have been described in detail already.  SMC uses a
frozen
polarised deuteron target [3], SLAC experiment E142 uses a polarised $^3He$ gas
target [4].  To the extent that $^3He$ is dominantly in an S-state such that
its
  two
protons couple to net spin zero (by the Pauli principle), the $^3He$
polarisatio
 n
directly leads to neutron polarisation.  There will certainly be debate about
th
 e
effect
of nuclear binding on the interpretation of data and extraction of the nucleon
structure functions (indeed this has already begun [5]).  In this talk I shall
a
 ssume
that
the experiments have satisfactorily taken this into account.

The material is organised as follows.

I shall begin by recalling the Genesis of the excitement, namely the proton
spin

measurement by EMC.  We shall see that both the original measured value for
the integral
$$
I^p \equiv \int dx g^p_1(x)
$$
of the polarised structure function has been modified as has the theoretical
prediction for its magnitude.  The outcome is that the spin ``crisis" has been
much reduced and is now only a 2$\sigma$ effect (for the proton at least).

Most attention has focused on the magnitude of the integral which sums over
valence and sea collectively and may even be affected by gluon polarisation.
In

``Exodus" I shall leave the integral and concentrate instead on the $x$-
dependence of the polarisation asymmetry
$$
A(x,Q^2) = g_1(x,Q^2)/F_1(x,Q^2)
$$
This is the quantity which is most directly measured and for which predictions
were made within the framework of a naive valence quark model more than 20
years ago.  We shall see that the data on both proton and, now, neutron are in
remarkable agreement with prediction throughout the valence quark region
$(x\gapproxeq$ 0.1).  The message, for me at least, is that the valence quarks
a
 re
polarised as ``expected"; hence {\bf if} the total integral $I^p$ violates the
theoretical expectation, this implies that the sea $(\bar{q}$ and/or gluons)
has
  a
non-trivial role.

In ``Numbers" I review the theoretical expectations for the magnitudes of the
integrals.  These depend on the F/D parameter for the octet of beta-decays and
I

present an updated value for this quantity and, in turn, for the sum rules.

I shall then return to the question of the integrals and in ``Deuteronomy"
shall

compare the new information from SMC (deuteron) and E142 ($^3He$, in effect
neutron).  Combining these results with the original EMC proton data reinforces
the message that care is needed in interpretation of these data and that there
i
 s
less of a ``crisis" than advertised hitherto.

In ``Revelations" I show that when all data sets are analysed at the same $Q^2$
(11 GeV$^2$) and with the modern {\bf un}polarised $F_1(x)$, a consistent
picture emerges.  The result is that the net quark spin is about 50\% of that
measured at low $Q^2$, and in accord with expectation from QCD evolution.  It
implies that the sea is polarised and I conclude by discussing  a possible way
o
 f
probing a polarised sea by
means of inclusive $K^-$ production at HERA (HERMES collaboration).

\section{Genesis}

Most interest has centred on the integral
$$
I^{p(n)} \equiv \int dx g_1^{p(n)} (x)
$$
of the polarised structure function.  There is an important sum rule due to
Bjorken [ 6] which follows rigorously from QCD.  Its expected value varies
slightly with
momentum transfer $Q^2$.

$$
I^p-I^n = 0.210 (1-\frac{\alpha (Q^2)}{\pi}) \matrix{Q^2=11 GeV^2\cr
\longrightarrow\cr} 0.192
$$
where I have specialised to $Q^2 = 11 GeV^2$ as a suitable reference value.
(Ref  [7]
has evolved all data to a common value in order to make a unified comparison).

Polarisation measurements from both protons and neutrons are needed to test
this
fundamental sum rule and, until recently, only proton data were available.
Gilman [ 8] had shown how to write sum rules for $p,n$ separately in terms of
an

(unknown)
SU(3) singlet contribution.  This was developed by Ellis and Jaffe [9] who made
the
assumption that the strange sea within the proton and neutron is unpolarised,
and
predicted $I^{p,n}$.  Their sum rule was expressed as a function of the F/D
parameter
that summarises nucleon and hyperon beta-decays.

The sum rule for the proton was written
$$
I^p = \frac{1}{12}(\frac{g_A}{g_V}) [1+\frac{5}{3} \frac{3F-D}{F+D}]
$$
and if F/D = 0.63 $\pm$ 0.02 (as assumed in ref 1, 2) the theoretical value
becomes 0.20 (reduced to 0.191 after $0(\alpha_s$) QCD corrections are
included)
 .
The original measurement of $I^p$ by EMC quoted
$$
I^p \mbox{(EMC, ref 1)} = 0.116 \pm 0.012 \pm 0.026
$$
and was compared with
$$
I^p \mbox{(TH)} = 0.191  \pm 0.002
$$

This shortfall of 40\% in the integral translated into the astonishing
conclusion that the net quark spin
$$
\Delta q \simeq 0
$$
Thus was the so-called proton ``spin crisis" born.

It is important to note that {\bf both experimental and theoretical values have
changed relative to the original numbers} quoted above.  The modern values are
$$
I^p \mbox{(EMC, ref 2)} = 0.126 \pm 0.011 \pm 0.014
$$
while the theoretical expectation is more appropriately
$$
I^p (\rm TH) = 0.175 \pm 0.007
$$
leading to a much reduced significance for the discrepancy.

I would first like to bring up to date the experimental situation concerning
$I^
 p$
and the status of criticisms that Roberts and I made at the time [10].  We made
three particular criticisms; namely the sensitivity to\\
i) F/D\\
ii) extrapolation as $x\rightarrow0$\\
iii) unpolarised $F_1(x)$ used to obtain $g_1(x)$ from $A(x)$.

Point (i):  I shall review the status of F/D in
section (3).

Point (ii):  we considered both a hypothetical $[x\ln^2(x)]^{-1}$  and
$x^\alpha
 $
form.  The former, it is now generally agreed, is not well
motivated and is withdrawn; the latter    (Regge) power fit for $x\lapproxeq$
0
 .1
has been used widely and contributes typically
$I$(0;0.01) $\simeq$ 0.002.  In view of our deepening understanding of the
$x\rightarrow 0$ behaviour
of $F_1(x)$ this may merit re-evaluation in future but I shall continue to
adopt

this  conventional form
here.

Point (iii)  The choice of unpolarised structure functions
in constructing  $g_1(x)$ is a significant concern and there have been
considerable
developments here since the original EMC data appeared. To help estimate these
uncertainties it is useful to recall the reasons why the quoted $I^p$ has
change
 d.

The contribution in the measured region ($0.01<x<0.7)$ was originally computed
using $F(x)$
of EMC.  This gave
$$
I(0.01;0.7) = 0.113 \pm 0.012
$$
(at $Q^2$ = 10.7 GeV$^2$).
The contributions in the unmeasured regions were estimated to be
$$
I(0;0.01) \simeq 0.002 \quad ; \quad I(0.7,1) \simeq 0.001
$$
and hence the original reported value [1]
$$
I^p = 0.116 \pm 0.012 \pm 0.026
$$
In their subsequent paper EMC  made a detailed comparison of how the integral
varied when  other $F_1(x)$
structure function data were used, notably those of BCDMS;   this gives [2]
$$
I^{BCDMS} (0.01;0.7)= 0.127\pm 0.014
$$
Using an average of $I^{EMC}$ and $I^{BCDMS}$ led to the total integral [2]
$$
I^p (EMC+BCDMS) = 0.123\pm 0.013\pm 0.019
$$

Independent of the above SLAC had measured their own $I^p$.  The
combination of the EMC(BCDMS) and SLAC raised $I^p$ slightly and reduced
the
errors, giving the modern value [2]
$$
I^p = 0.126\pm 0.010\pm 0.015
$$

There are two developments that imply that reevaluation of this value is
warranted.   First, we now have improved understanding of the relative
normalisations
of the various $F_1(x)$ (this is reviewed in ref 11).  In ref [10] we used
BCDMS

data, hence our $I^p\simeq$ 0.14.
The NMC data at $x<0.1$ (which link smoothly onto the
$x<0.01$ data from HERA) suggest that the $F_1(0.01<x<0.1)$ may be rather
bigger then thought hitherto.  This in turn would cause $g_1(x)$ to increase in
this region with corresponding increase in the extrapolated $I$(0,0.01).  Thus
a

new computation of $g_1(x)$ from the measured $A(x)$ is called for. A
preliminary investigation [7] suggests $I^p\simeq$ 0.138.

I shall critically examine the theoretical prediction for the integrals later.
 First I
shall concentrate on the directly measured asymmetry by means of a pedagogical
example.

\section{X-odus}

Most analyses of the ``spin crisis" have concentrated on the {\bf integral} of
$g_1(x)$.
Here I shall concentrate on the detailed dependence of the {\bf asymmetry} in
the
valence region.  The {\bf {\it x} dependence}  will provide important clues as
t
 o
the
source of the ``spin crisis".

\subsection{A simple pedagogic example and its generalisation}

I will begin with the constituent quark model in order to give a specific
exampl
 e
of
more general ideas that will be developed later.  Although this model has no a
priori justification in the case of deep inelastic polarised structure
functions
 ,
nonetheless it will be seen to give a remarkably good average (in $x_{bj}$)
description of some of the data.  Thus it may provide a suitable reference from
which intuition may be developed and more legitimate models built.

Given that all pairs of coloured quarks attract into $\bar{3}$, which are
antisymmetric under interchange, it follows from Pauli that the pairs must be
symmetric under interchange of all other quantum numbers.  Let us suppose
that
they are symmetric in momentum space (this is the case in the constituent
picture
where the quarks are in overall $S$-wave).  This implies that identical
flavours

(such as $uu$ in the proton or $dd$ in the neutron) couple symmetrically in
spin

(thus to $S=1$ in the nucleon rest frame).

Coupling the $S=1$ with the $S=1/2$ of the remaining valence quark to form
overall $J$=1/2 (this is the rest frame constituent quark description) yields
\be
p^\uparrow = \sqrt{\frac{2}{3}} (u^\uparrow u^\uparrow) d^\downarrow +
\sqrt{\frac{1}{3}} (u^\uparrow u^\downarrow) d^\uparrow
\ee
where arrows refer to the $\hat{z}$ component of spin.  This implies that the
spin
weighted numbers of flavours are
\be
u^\uparrow  = 5/3 \quad u^\downarrow =1/3 \quad d^\uparrow = 1/3 \quad
d^\downarrow = 2/3.
\ee
and so
\be
\begin{array}{llll}
u & \equiv u^\uparrow + u^\downarrow & \equiv \int dxu(x) &= 2\\
d & \equiv d^\uparrow + d^\downarrow & \equiv \int dxd(x) &= 1\\
\Delta u & \equiv u^\uparrow - u^\downarrow & \equiv \int dx\Delta u(x) &
=
4/3\\
\Delta d & \equiv d^\uparrow - d^\downarrow & \equiv \int dx\Delta d(x) &
= -
1/3
\end{array}
\ee

This model clearly has all of the proton's spin carried by its valence quarks
\be
\Delta q\equiv \Delta u + \Delta d = 1
\ee
and immediately demonstrates the ``bad" SU(6) result for $g_A/g_V$ since
\be
\frac{g_A}{g_V} \equiv \Delta u - \Delta d = \frac{5}{3}
\ee
Closer examination of the derivation of this result suggests that it should
read
  in
general (see e.g. ref  13)
\be
\frac{g_A}{g_V} =\frac{5}{3} \Delta q \longrightarrow \Delta q \simeq 0.75
\ee

To the extent that nucleon magnetic moments probe the spin magnetic
moments
of the constituents and assuming that the latter are proportional to the
electri
 cal
charges of
the quarks, then the ratio $\mu_p/\mu_n$ is in proportion to $\Delta u
/\Delta
d$
\be
\frac{\mu_p}{\mu_n} = \frac{\frac{2}{3}\Delta u - \frac{1}{3}\Delta
d}{\frac{2}{3}\Delta d - \frac{1}{3}\Delta u} = \frac{2(\frac{\Delta u}{\Delta
d
 })-
1}{2-(\frac{\Delta u}{\Delta d})}
\ee
(where $\Delta d^n \equiv \Delta u^p\equiv \Delta u$ and $\Delta
u^n\equiv\Delta d^p \equiv \Delta d$ in the above).  We see that the empirical
result follows successfully
\be
\frac{\mu_p}{\mu_n} = -\frac{3}{2} \leftrightarrow \frac{\Delta u}{\Delta d} =
-4.
\ee

Now consider the $<e^2_i>$ weighted $\Delta  q_i$.  This is intimately related
to
what is probed in spin dependent inelastic scattering of leptons on polarised
nucleons in the $x\gapproxeq$ 0.1 region dominated by valence quarks [13].

This ``asymmetry"
\be
A = \Sigma e^2_i \Delta q_i /\Sigma e^2_i  q_i.
\ee
For a neutron
\be
A^n \simeq \frac{1}{9} \Delta u + \frac{4}{9} \Delta d \rightarrow 0
\ee
which follows because $\Delta u= - 4 \Delta d$.  For a proton
\be
A^p = \frac{4}{9} \Delta u + \frac{1}{9} \Delta  d \equiv
\frac{1}{6} (\Delta  u - \Delta d) + \frac{5}{18}   (\Delta  u + \Delta
d)\rightarrow \frac{5}{9}
\ee

For future reference note that
\be
A^n ( \Sigma e^2_i q_i)^n = \frac{5}{18}
\Delta q - \frac{1}{6} \frac{g_A}{g_V}
\ee
\be
A^p ( \Sigma e^2_i q_i)^p = \frac{5}{18}
\Delta q + \frac{1}{6} \frac{g_A}{g_V}
\ee

The above picture would apply to the deep inelastic polarisation in a primitive
picture where the nucleon is made of three quarks each of which, symmetrically,
carries fraction $x = 1/3$ of the parent's momentum.  Thus the structure
function
$\sim \delta (x-1/3)$.  In this toy model, the above results may be written
\bea
2F^n_1(x) = \frac{2}{3} \delta (x-1/3) & ; & A^n(x) = 0\nonumber\\
2F^p_1(x) = \delta (x-1/3) & ; & A^p(x) = \frac{1}{3} \frac{g_A}{g_V} \delta
(x-
1/3)
\eea
where $F_1$ is the unpolarised transverse structure function of the target
(note

that
$2xF_1(x)\equiv F_2(x) = \Sigma e^2xq(x)$).

It is well known that the ratio $\frac{F^n_1}{F^p_1} (x=1/3) \simeq
\frac{2}{3}
 $
in
the real world; thus it was suggested over twenty years ago, in advance of any
data
for polarised asymmetries, that the $A^{n,p}$ predictions may also be realised
when
$x\simeq 1/3$.  Thus it is quite remarkable that the $A^p (x=1/3) $, measured
in

1977 and 1987, and the new reports of $A^n$ (from either polarised $^3He$ or
deuteron)  are in complete agreement with these predictions, (fig 1).

In 1972 Gilman suggested that eqn 14 be generalised in an integrated form for
non-
diffractive (valence quark) contributions, thus
\be
\int dx (F^p_1(x) - F^n_1 (x)) = 1/6
\ee
\be
\int dx (g^p_1(x) - g^n_1 (x)) = 1/6  \frac{g_A}{g_V}
\ee
where $g_1(x) \equiv  F_1(x) A(x)$.  Thus one sees the origin of the
Gottfried (eqn 15) and Bjorken (eqn 16) sum rules in this toy model [13].  But
i
 t
also,
through eqns (12, 13), leads to sum rules for the proton and neutron
individuall
 y
\be
\int dx g_1^{(\stackrel{p}{n})}(x) =  \pm \frac{1}{12} \frac{g_A}{g_V}
+\frac{5}{36}\Delta q
\ee
These correspond to the Ellis-Jaffe sum rules in the particular case where $F/D
 =
2/3$.  For general $F/D$ one may write
\be
\frac{g_A}{g_V}  \equiv F+D \; ;
\ee
\be
\Delta q \equiv 3F-D
\ee
and so eq (17) generalises to the Ellis-Jaffe form
\be
\int dx g_1(x) = \frac{1}{12} \frac{g_A}{g_V} [\pm 1 + \frac{5}{3} \frac{3F-
D}{F+D}]
\ee

The above picture at $\delta(x-1/3$) was ``generalised" to an integrated form
which is the Ellis-Jaffe sum rule.  {\bf If} the latter is violated by data,
per
 haps we
can use intuition from the above example to see where the violation originates.

In particular we saw how the sum rule is derived from the measured
asymmetry, $A(x)$, albeit somewhat remotely.  The $A(x\simeq$ 1/3) predicted
in the simple model is remarkably good.  Let us return to that model and see
what {\bf pre}dictions it made for $A(x\gapproxeq$ 0.1) {\bf several years}
prio
 r
to any data on polarised deep inelastic scattering.

Guided by the unpolarised data, where $\frac{F_1^n}{F^p_1} (x\rightarrow
1)<\frac{2}{3}\rightarrow \frac{1}{4}$ I suggested that $A^{p,n}(x\rightarrow
1)\rightarrow 1$.  The foundations for this are described in some detail in
chapter 13 of ref ( 13 and I shall not repeat them here.  The latter prediction
 was
put on more solid ground following the advent of QCD and the development of
counting rules [14].  The latter imply  [15] that for a system of $N$
constituen
 ts
$$
F_{a/A} (x) \sim (1-x)^{2N-1+2|\Delta h|}
$$
where $\Delta h$ is the difference in helicity
$$
\Delta h = |h_a-h_A|
$$
and hence that $A(x\rightarrow 1)\rightarrow 1$.  Prior to these helicity
dependent counting rules I
had advocated [12]
$$
A^n(x\rightarrow 1) = A^p(x\rightarrow 1) =\xi
$$
where $\xi$ = fraction of hadron spin carried by quarks.  Assuming this to be
75\% generated the curves in ref (12).  Specifically if $F^n_1/F^p_1(x) \equiv
R(x)$ then
$$
A^p(x) = \frac{19-16R(x)}{15} \xi
$$
$$
A^n(x) = \frac{2-3R(x)}{5R} \xi
$$
These curves are compared with the subsequent proton and neutron data in fig
2.  The agreement in the valence region is quite remarkable and suggests
that {\bf the polarisation response of valence quarks at large $Q^2$ is rather
similar to
that anticipated from our low energy experience}.  Why this should be so is
perhaps an interesting question in its own right, but it does suggest that any
problems in the integrated structure function do not originate here.  We must
focus either on the sea polarisation (of either gluons or antiquarks) and/or on
the strength of the unpolarised $F_1(x)$ by which the $A(x)$ is multiplied in
order to reach $g_1(x)$.

In order to help determine what is the message of the sum rules, we must first
establish what the sum rules are.  The literature has not always been
consistent

on this.

\section{Numbers}

In order to define the sum rules, first write
$$
I^p=I_3+I_8 + I_0
$$
$$
I^n=- I_3 +I_8 +I_0
$$
where [16]
\bea
I_3 = \frac{1}{12} (1-\frac{\alpha_s}{\pi}) a_3\nonumber \\
I_8 = \frac{1}{36} (1-\frac{\alpha_s}{\pi}) a_8\nonumber \\
I_0 = \frac{1}{9} [1-(1-\frac{2N_f}{\beta_0} )\frac{\alpha_s}{\pi})
a_0\nonumber
\eea
In terms of quark polarisation $\Delta q\equiv \int dx (q^\uparrow (x)
-q\downarrow (x)$) , or the F and D parameters,
$$
\begin{array}{lll}
a_3 & = \Delta u - \Delta d & \equiv F+D\\
a_8 & = \Delta u + \Delta d-2\Delta s & \equiv 3F-D\\
a_0 & = \Delta u + \Delta d+\Delta s
\end{array}
$$
so that $a_0$ corresponds to the net quark polarisation.  Then since
$F+D\equiv\frac{g_A}{g_V}$ we have the Bjorken sum rule
$$
I^p-I^n = \frac{1}{6} (\frac{g_A}{g_V}) (1-\frac{\alpha_s}{\pi})\rightarrow
\left\{ \matrix{0.210& \alpha\rightarrow 0\cr
0.192 & \alpha \simeq 0.25\cr}\right.
$$
For the separate nucleons, and specialising to $N_f$ = 3
$$
I^p_n = \frac{1}{12} (\frac{g_A}{g_V}) \left\{ [
\pm 1+\frac{5}{3} (\frac{3F-D}{F+D})] - \frac{\alpha_s}{\pi}[\pm 1+\frac{7}{9}
(\frac{3F-D}{F+D})] \right\} +\frac{\Delta s}{3} (1-\frac{\alpha_s}{3\pi})
$$
which reduces to the Ellis-Jaffe sum rule when $\alpha_s \rightarrow 0$ and if
$\Delta s=0$.  Note that $(\frac{g_A}{g_V}) $ at the front and (F+D) in
denominator are identical quantities: this has not always been maintained in
analyses.

For the case of 3 flavours we may rewrite them
in the form
$$
I^p = \frac{1}{6} (F+\frac{1}{3} D) (1-\frac{\alpha}{\pi}) + \frac{1}{9} \Delta
 q (1-
\frac{\alpha}{3\pi})
$$
$$
I^n = \frac{1}{9}  D (1-\frac{\alpha}{\pi}) + \frac{1}{9} \Delta q (1-
\frac{\alpha}{3\pi})
$$
$$
\frac{1}{2} I^{p+n}  = \frac{1}{12} (F-\frac{1}{3} D) (1-\frac{\alpha}{\pi}) +
\frac{1}{9} \Delta q (1-\frac{\alpha}{3\pi})
$$
Note that $I^{p,n}$ do {\bf not} share a common factor of  $(1-
\frac{\alpha}{\pi})$.  (Contrast the expressions at eq 3 in ref 4 and ref (17)
f
 rom
which it originates.
 The analysis in ref 4 is thereby
compromised).  Furthermore E142 use F = 0.47 $\pm$ 0.04  D = 0.81 $\pm$ 0.03
from Jaffe-Manohar
analysis [17],
whose sum does not centre on  the currently accepted value of
$(g_A/g_V)_{np}$:
this introduces further
incongruity.  As these F, D values [17]  are determined from 1986 data, which
ar
 e
now
significantly modified, a modern fit is called for.

I list the current world averages from the 1992 edition of the Particle Data
Gro
 up
[18],
together with their F, D parameterisation and the SU(6) value (arbitrarily)
renormalised such that only 25\% of the spin is ``lost" to relativistic (lower
components of spinors, see ref 12 and 13).

$$
\begin{array}{l|l|l|l}
g_A/g_V & F, D & {\rm experiment} [18] & \frac{3}{4} (SU_6)\\\hline
np & F+D & 1.2573\pm 0.0028 & 5/4\\
\Lambda p & F+\frac{1}{3}D & 0.718\pm 0.015 & 3/4\\
\Xi\Lambda & F-\frac{1}{3}D & 0.25 \pm 0.05 & 1/4\\
\Sigma n & F-D & -0.340\pm 0.017 & -1/4\\\hline
\end{array}
$$
The best fit, with $\chi^2$ = 1.55 for one degree of freedom (F/D with F+D
constrained to equal 1.257) is [7]
$$
\left. \begin{array}{lll}
F& = & 0.459\pm 0.008\\
 D &=& 0.798\mp 0.008\end{array}
\right\} F/D
= 0.575 \pm 0.016
$$
This is 1$\sigma$ larger than in Close-Roberts 1988 paper  [10] principally due
 to
improved $\Lambda p$ and $\Sigma n$ data.  I shall use these values in what
follows; however there are two caveats.  First there is a systematic error, not
included, whch arises from the phase space or form factor corrections in the
$\Delta S$ = 1 examples [19].  The second is potentially more serious.

The quoted figures assume that in the hadronic axial current
$$
A_\mu = g_A\gamma_\mu \gamma_5 - g_2 \frac{i\sigma_{\mu\nu}q^\nu
\gamma_5}{m_i+m_j}
$$
one has $g_2$ = 0.  While this is assured in the limit where $m_i=m_j$ (such as
$n\rightarrow p$) it is not necessarily so for $\Delta S=1$.  Indeed, in quark
models one expects that $g_2=0 (\frac{m_i-m_j}{m_{ij}}$) with $m_{ij}\equiv
\frac{1}{2}(m_i+m_j)$. [20]

The Hsueh et al. analysis of $\Sigma n$ made a fit allowing for $g_2\neq 0$ and
found [21]
$$
g_2 = -0.56 \pm 0.37
$$
$$
g_A = 0.20 \pm 0.08
$$
This raises a tantalising possibility that the $(g_A/g_V$) throughout the octet
 are
given by the naive quark model, all values being renormalised by 25\% (such
that the net spin is 0.75 rather than 1).  This is illustrated in the fourth
col
 umn of
the above table.  Such an eventuality would correspond to the realisation of
the

effective quark model result
$$
F= 1/2, \; D = 3/4 \quad ; \quad F/D = 2/3
$$

We note that either of these solutions gives essentially the same predictions
fo
 r
the non-singlet contributions to $I^{p,n}$\,
$$
\frac{1}{6} (F + \frac{1}{3} D) = 0.131 \; versus \; 0.125
$$
$$
\frac{1}{9} D = 0.09 \; versus \; 0.08
$$
So the inferred $\Delta q$ from $I^p$ or $I^n$ changes only marginally; the
sensitivity is $\Delta s$  [22] since
$$
\Delta s \simeq  -(F-\frac{D}{3}) +\frac{1}{3} \Delta q
$$
and
$$
F-\frac{1}{3} D = 0.19 \; versus \; 0.25
$$
It is the assumption that $\Delta s$ = 0 in the Ellis-Jaffe sum rule and the
sensitivity to the combination $F-\frac{1}{3}D$ that causes the sensitivity of
t
 he
$I^{p,n}$ (THY) to F,D.

Using the best fit F, D in the standard $g_2$=0 approach, the predictions
are
\bea
I^p & = & (0.121\pm 0.001)(1-\frac{\alpha}{\pi}) + \frac{1}{9} \Delta q (1-
\frac{\alpha}{3\pi})\nonumber\\
I^n & = & (-0.089\pm 0.001)(1-\frac{\alpha}{\pi}) + \frac{1}{9} \Delta q (1-
\frac{\alpha}{3\pi})\nonumber\\
\frac{1}{2} I^{p+n} & = & (0.016\pm 0.001)(1-\frac{\alpha}{\pi}) + \frac{1}{9}
\Delta q (1-\frac{\alpha}{3\pi})\nonumber\\
\eea
We are now in position to confront these with the new data.

\section{Deuteronomy}

SMC have measured the deuteron (essentially the sum of proton and neutron).
Roberts and I [10] had pointed out that this has direct interest in that the
$I^{p+n}$ is less sensitive to errors than $I^p$ or $I^n$ separately if one
wish
 es to
extract the net quark spin.

The sum rule for the proton  immediately highlights a sensitivity in the $I^p$
data.  To
the extent that
$I^p(EMC)\simeq$ 0.12 then $\Delta q\rightarrow 0$ if $\alpha_s\rightarrow
0$, which is the much heralded spin crisis.  However, the $\frac{1}{9}$ in
front
  of
$\Delta q$  causes a rapid turn on of the inferred $\Delta q$ when any small
changes are included in $I^p$ or $\alpha_s$.  For example
$$
\Delta q \simeq \frac{9\alpha_s}{\pi} 0.12 + 9(I^p - 0.12)
(1+\frac{\alpha}{3\pi
 })
$$
The first term alone generates 10\% spin; systematic and statistical errors in
$I^p$ each induce large uncertainties in $\Delta q$.  This is what underlies
the

result that the deviation is only a 2$\sigma$ effect.  Moreover, higher order
corrections in $0(\alpha_s)$ at finite $Q^2$ may also induce corresponding
ninefold increases in the corrections to $\Delta q$.  Misestimates of the
contribution to $I^p$ as $x\rightarrow 0$ will also affect the inferred value
of

$\Delta q$; we shall see later that deuteron data may enable this particular
problem to be circumvented.

$I^p$ and $I^n$, are expressed in general as a linear sum of $g_A/g_V$ and
$\Delta
q$.  Thus when extracting $\Delta q$ from $I^p$ (or $I^n$) one must first
remove

the ``unwanted" contribution  ($g_A/g_V$).  This is what, in part, causes
small percentage errors in $I^p$ to be larger percentagewise in 
$\Delta q \simeq 9(I^p-(g_A/g_V))$ and, furthermore, the factor of 9 makes 
things worse still.

It is this sensitivity that was the main message in ref [10], being more
general

than
the particular example used as illustration there.  In particular, the
fluctuati
 on in
the three data points at $x\rightarrow 0$ in the EMC proton data could arguably
take care of (some of) the 2.5$\sigma$ effect if future data shows that the
asymmetry values given by these data points are at the upper end of their
magnitudes. (Note that the NMC unpolarised structure function [23] tends to
suggest that the $g_1(x)$ is underestimated here [7].)

We can form combinations of $I^{p,n}$ that emphasise the $\Delta q$ or
eliminate it.  Thus two natural combinations are
$$
I^p-I^n = 2I_3
$$
$$
I^p+I^n = I_8+I_0
$$
The former has eliminated $\Delta q$  (and is the Bjorken sum rule
combination).  In ref 10 we  noted that the orthogonal combination
(essentially
given by $I^d$) emphasises $\Delta q$.

First recall the individual nucleon integrals
$$
I^p_n \simeq  (\matrix{0.12\cr -0.09\cr})  (\frac{g_A}{g_V})
 (1-\frac{\alpha_s}{\pi}) + \frac{\Delta q}{9}  (1-\frac{\alpha}{3\pi})
$$
The coefficient of $(\frac{g_A}{g_V})$ saturated the $I^p$.  But now form
$I^N=\frac{1}{2}(I^p+I^n$)  which to good approximation may be measured
directly with a deuteron target,
$$
I^N \simeq  (0.016)   (\frac{g_A}{g_V})    (1-\frac{\alpha_s}{\pi}) +
\frac{\Delta q}{9} (1-\frac{\alpha}{3\pi})
$$
We see that the ``unwanted" $\frac{g_A}{g_V}$ term has been suppressed so
that $\Delta q$ has a chance of showing up.  At first sight one may say that
the
 re
is no free lunch and that $I^N\simeq 0$ so that one still has problems.  This
is

indeed true for the total integral; however one may place an {\bf upper bound}
on $\Delta q$ (and hence if this is small one has established the spin crisis).

The reason lies in the prediction (now confirmed) for the behaviour of $A(x)$,
and hence $g_1(x)$, in the valence region, $x\gapproxeq 0.1$.  It is known that
$A^p(x)>0$, and was predicted  [12, 24] that $A^n(x)>0$ certainly for
$x\gapproxeq$ 0.3.
The $A^n(x\rightarrow 0)\lapproxeq$ 0 and so theoretical prejudice (and maybe
even data) suggest that
$$
I^d (x<x_c)<0\quad ;\quad I^d(1>x>x_c)>0
$$
for some small $x_c$.
If we assume that $I^d(x\rightarrow 0$) does not oscillate,
one can immediately measure an upper bound to $\Delta q$.
$$
\int^1_{x_c} g_1^d(x) dx \geq \int^1_0 g_1^d(x) dx \simeq
0.016  (\frac{g_A}{g_V}) (1-\frac{\alpha_s}{\pi}) +\frac{1}{9} \Delta q (1-
\frac{\alpha}{3\pi})
$$
We will now confront this with data and consider the implications of the new
experiments.  First we shall compare the experiments while ignoring the fact
tha
 t
they span different ranges of $Q^2$.  This is done merely for first orientation
 and
to highlight the importance of careful treatment of $Q^2$: in the next section
w
 e
shall see how a comparison of the experiments at a {\bf common} $Q^2$ value
leads to rather different conclusions.

The SMC measurement on the deuteron gives [3]
$$
I^n +I^p = 0.049 \pm 0.044 \pm 0.032
$$
which on combining with $I^p$ of EMC yields
$$
I^n = - 0.08 \pm 0.04 \pm 0.04.
$$
In turn this would give
$$
I^p-I^n = 0.206 \pm 0.06
$$
which is compatible with the Bjorken sum rule (though with large
uncertainties).

The E142 data are at rather small $Q^2$ and give for the neutron ``directly"
$$
I^n = - 0.022 \pm 0.011
$$
If this were the only datum available we would compare with the Ellis-Jaffe sum
rule
and conclude that all is well with the world.  (Though, as we shall see in
secti
 on
5, the small value of $Q^2$  require this to be re-evaluated.)   If now one
combines this result
with
the $I^p$ from EMC one obtains
$$
(I^p-I^n) (EMC \; \mbox{and}\; E142) = 0.148 \pm 0.022
$$
which is about two standard deviations below the fundamental Bjorken sum
rule
(this 2$\sigma$ reflecting the 2$\sigma$ shortfall of $I^p$ from EMC relative
to

Ellis-Jaffe).

As a final way of combining the three experiments pairwise we can take $I^n$
from
E142 and $I^{p+n}$ from SMC.  This would imply
$$
I^p = 0.071 \pm 0.055
$$
and hence
$$
I^p -I^n = 0.093\pm 0.08
$$

We can now take combinations of all experiments and compare their quoted
integrals
pairwise with the $I^{p,n}$ and $I^{p-n}$.  This gives
$$
\begin{array}{l|l|l|l|l}
& I^p & I^n & I^d &I^{p-n}\\\hline
{\rm Theory} & .173 & -.019 & .15 & .192\\
& (.015)& (.011) & (.02)\\\hline
EMC+SMC & .13 & -.08 &&.21\\
& (.02) & (.06) && (.06)\\\hline
E142+EMC && -.02 & .10 & .15\\
&&(.01) & (.03) & (.02)\\\hline
SMC+E142 & .07 && .05 & .09\\
& (.06) && (.05) & (.08)\\\hline
\end{array}
$$
Note that the ``theory" value corresponds to $Q^2$ = 11 GeV$^2$ whereas the
experiments span a range of $Q^2$.
Superficially only the first of these appears to survive ``self consistency"
che
 cks:
proton and neutron each violate the E-J sum rule in a common manner that
preserves the fundamental Bjorken sum rule.  The second possibility violates
Bjorken due to some anomaly that is manifested in proton data alone.
The third combination also appears to have an equivalent
``inconsistency" to the above and one may be tempted therefore to ``blame" the
SLAC neutron data; however, the neutron data satisfy  the E-J sum rule and
would appear ``innocent" if that were all that we knew.

However this comparison is misleading because we have cavalierly compared
and combined data from experiments at
different $Q^2$ and whose $g_1(x)$ have been constructed from the $A(x)$ by
invoking different data on $F_1(x)$.  In ref 7 we have attempted to compare the
experiments under common assumptions.   This has an interesting consequence
which I now summarise.

\section*{Revelations}

The sensitivity to the unpolarised structure function, and $Q^2$ dependence, is
highlighted by the following example [7].  Using the NMC data for $F_1^p(x)$
the

EMC proton integral and inferred $\Delta q$ become
$$
I^p = 0.134 \pm  0.012\quad ; \quad \Delta^pq = 22.9 \pm 12.0\%
$$
For the E142 neutron, evolution of the data to a common value $Q^2$ = 11
GeV$^2$ yields (assuming that the asymmetry is $Q^2$ independent)
$$
I^n = -0.031 \pm  0.007\quad ; \quad \Delta q = 49.8 \pm 6.7\%
$$
The Bjorken sum rule is to be compared with
$$
I^{p-n} = 0.165 \pm 0.014
$$
There is  a discrepancy between the $I^d$ of SMC and that constructed from
the $p+n$ of EMC and E142.  When one examines the $g_1(x)$ directly one
notices, in particular, that the $x\gapproxeq $ 0.4 data of SMC have large
error
 s
and give a lack of strength in this region relative to the $p+n$ combination.
W
 e
suggest [7] that the $p+n$ ``constructed data" be used as the guide here
because
 \\
a) they are consistent with the SMC and have smaller errors\\
b) the $A^d(x\rightarrow 1$) is predicted to be large (unity ?) as
$x\rightarrow
  1$
and this also favours the high end of the SMC error bars, consistent with the
EMC + E142 combination.

Taking this as the most probable solution in the $x\gapproxeq 0.4$ valence
quark

region, the constructed $I^d$ and resulting $\Delta q$ are
$$
I^d = 0.037 \pm 0.018\quad; \quad  \Delta q = 21.6 \pm 17.8%
$$
In fig (3) I summarise the various integrals and $\Delta q$ from these
``improved" data.  It is particularly notable that {\bf all combinations now
giv
 e
excellent agreement with the fundamental Bjorken sum rule}.  (Note that at
small $Q^2$ higher order corrections must be included such that [29]
 for three flavours
$$
I^{p-n} \simeq
\frac{1}{6} \frac{g_A}{g_V} (1-\frac{\alpha}{\pi} - 3.6
(\frac{\alpha}{\pi})^2-20 (\frac{\alpha}{\pi})^3)
$$
If  $\frac{\alpha}{\pi}$ = 0.1 this reduces $I^{p-n}$ to some 0.177.  Similar
ef
 fects
may be expected for $I^{p,n}$ separately.  Higher twist can also be important,
especially in the case of the neutron [30,31]; $\delta I^n\simeq$ 0.04/$Q^2$,
$\delta I^p\simeq$- 0.004/$Q^2$.  Ref [30] has shown that these can make a
significant effect on the analysis, especially as the neutron E142 data have
low

$Q^2\simeq$ 2 GeV$^2$  and conclude that
$$
\Delta q = 22 \pm 10\%
$$
This is in accord with our results abstracted from $d$ and $p$ experiments
above

(22 $\pm$ 18\% and 23 $\pm$ 12\% respectively) as is to be expected since these
tended to have $<Q^2>$ large and hence be less susceptible to $1/Q^2$
corrections.   However the value inferred for $\Delta q$ depends rather
sensitively on the assumed values of the higher twist coefficients.  The above
value is valid to the extent that one accepts the QCD sum rule input; if
however

one chooses to fit the $A/Q^2$ terms such that a common value of $\Delta q$
obtains for each of the three experiments then [7]  finds $\Delta q$ 38 $\pm$
48\%.
Clearly a deeper understanding of the magnitude of $1/Q^2$
corrections for $p$ and $n$ targets is required if $\Delta q$ is to be
extracted
  from
low $Q^2$ data.  Note again that
$$
\Delta q = 41 \pm 6\%
$$
is a minimum $\chi^2$ fit {\bf if} the neutron $(Q^2$ = 2 GeV$^2$) data are
included and evolved without any $1/Q^2$  higher twist contribution.

The conclusion seems to be that the nucleon spin structure as revealed in low
energy data is {\bf modified} when evolved to larger $Q^2$ but not totally
destroyed.  Higher twist and $0(\frac{\alpha}{\pi})^3$ corrections to the
Ellis-
Jaffe sum rules are probably needed before reliable conclusions on $\Delta q$
ca
 n
be drawn.  The relative spin strength of the $u$ and $d$ valence quarks is
preserved (as evidenced by the successful prediction of $A(x)$ in the valence
region) but some of their overall strength is evolved into the (polarised) sea
including glue and strangeness.

If the sea is polarised then we need experiments to reveal this directly.
Proce
 sses
that are dominated by gluons (such as $Q\bar{Q}$ production) may be tested for
target polarisation dependence in order to measure $\Delta G(x)$ (see refs 25).

The presence of polarised flavour in the sea [26] may be revealed by inclusive
leptoproduction of hadrons, in particular $K^-$.

{}From unpolarised deep-inelastic scattering the probability is estimated to be
greater than 70\% that the fastest forward-moving charged hadron with $z>$ 0.5
contains   a quark of the same flavour  as the scattered quark [27].

In principle the production of $K^-(s\bar{u}$) at large $z(=E_K/E_\gamma$) in
$\ell_p\rightarrow \ell^\prime K^-$ \ldots is most clear in this regard.  The
lepton interacts with a $q($ or $\bar{q}$) which subsequently produces the
detected hadron.  As $z\rightarrow 1$ the most probable occurrence is that the
hadron contains the struck $q(\bar{q}$) in its valence Fock state [13, 28];
thus
  a
fast $K^-$ is a signal for an $s$ or $\bar{u}$ having been struck by the
current
 .  As
$s, \bar{u}$ occur in the proton's sea, a target polarisation dependence of
fast

$K^-$ production would indicate that the $s$ or $\bar{u}$ components of the
proton sea are polarised; indeed the $K^-$ inclusive asymmetry as
$z\rightarrow 1$ is a direct measure of the amount and sign of the sea
polarisation.

This simple example is too idealised.  The limit $z\rightarrow 1$ is not
accessi
 ble
in practice and production of $K^-$ at $z<1$ is contaminated by the highly
polarised valence quarks.  However, in ref [26] we show that even for $z\leq$
0.
 5
there is a distinct dependence on sea polarisation.   Measurement of the $K^-$
dependence on beam-target polarisation promises to reveal polarisation in the
flavoured sea.  This experiment is planned to take place at HERA (the HERMES
Collaboration) within a few years.  This could clarify the way that a proton's
s
 pin
is distributed among its valence quarks and the flavoured $q\bar{q}$ sea.

\section*{Acknowledgements}

I am indebted to E Gabathuler, RG Roberts, S Rock, GG Ross and T Sloan for
discussions about the recent experiments.

\section*{Figure Captions}

\begin{itemize}
\item{ } $A^{p,n} (x=1/3)$ compared with prediction.
\item{ } Predictions for $A(x)$ if $\zeta$ = 1 (solid) or $\zeta$ = 3/4
(dashed)
  for
$p$ and $n$, compared with data.
\item{ } Diagrammatic representation of results and predictions.
\end{itemize}

\end{document}